\documentclass[a4paper,11pt]{article}
\pdfoutput=1 

\usepackage{jcappub} 

\def\apjl{ApJ}%
\def\apjs{ApJS}%
\def\physrep{Phys.\ Rep.}%
\usepackage{graphicx}
\usepackage{multirow}
\usepackage{graphicx}
\usepackage{url}
\usepackage{xspace}
\usepackage{tabularx}
\definecolor{darkblue}{rgb}{0,0,0.5}
\definecolor{darkgreen}{rgb}{0.1,0,0.3}
\definecolor{darkred}{rgb}{0.6,0,0}
\usepackage[normalem]{ulem}

\usepackage{geometry}      
\title{CNO Neutrino Grand Prix: The race to solve the solar metallicity problem}
 \author[a]{David G. Cerde\~no,}
 \affiliation[a]{Institute for Particle Physics Phenomenology (IPPP), Durham University, Durham DH1 3LE, United Kingdom} 
  \author[b]{Jonathan H. Davis,}
\affiliation[b]{Theoretical Particle Physics and Cosmology, Department of Physics,
King's College London, London WC2R 2LS, United Kingdom}
 \author[b]{Malcolm Fairbairn}
 \author[c,d]{and Aaron C. Vincent}
 \affiliation[c]{Department of Physics, Imperial College London,Blackett Laboratory, Prince Consort Road, London SW7 2AZ, United Kingdom}
\affiliation[d]{Canadian Particle Astrophysics Research Centre (CPARC), Department of Physics, Engineering Physics and Astronomy, Queen's University, Kingston ON K7L 3N6, Canada}
 
 \emailAdd{d.g.cerdeno@durham.ac.uk} 
  \emailAdd{jonathan.davis@kcl.ac.uk}
  \emailAdd{malcolm.fairbairn@kcl.ac.uk} 
 \emailAdd{aaron.vincent@queensu.ca}  
\abstract{Several next-generation experiments aim to make the first measurement of the neutrino flux from the Carbon-Nitrogen-Oxygen (CNO) solar fusion cycle.   We calculate how much time these experiments will need to run for in order to measure this flux with enough precision to tell us the metal content of the Sun's core, and thereby help to solve the solar metallicity problem.
For experiments looking at neutrino-electron scattering, we find that SNO+ will measure this CNO neutrino flux with enough precision after five years in its pure scintillator mode, provided its $^{210}$Bi background is measured to $1\%$ accuracy. By comparison, a 100~ton liquid argon experiment such as Argo will take ten years in Gran Sasso lab, or five years in SNOLAB or Jinping. Borexino could obtain this precision in ten years, but this projection is very sensitive to background assumptions. For experiments looking at neutrino-nucleus scattering, the best prospects are obtained for low-threshold solid state detectors   (employing either germanium or silicon). These would require new technologies to lower the experimental threshold close to detection of single electron-hole pairs, and exposures beyond those projected for next-generation dark matter detectors.}
\begin{document}
\maketitle
\flushbottom

\section{Introduction}
\label{sec:intro}

The Sun is powered by nuclear fusion reactions occurring in its core~\cite{Bethe:1939bt,Bahcall:2004pz,Bahcall:2000nu,Antonelli:2012qu,Davis:2016hil}. This involves hydrogen and helium predominantly, but also heavier elements such as lithium, carbon, nitrogen, oxygen, beryllium and boron. Each heavier element is produced through nuclear fusion from the lighter ones in a chain-reaction, with some steps in the chain releasing neutrinos of a characteristic energy spectrum. There are two main chains which convert hydrogen to helium in stars: the proton-proton chain and the carbon-nitrogen-oxygen (CNO) cycle~\cite{Bethe:1939bt}, both of which lead to neutrino production.
For the proton-proton chain there are five neutrino components (pp, $^7$Be, $^8$B, pep and hep) each with a different spectrum, while for the CNO cycle there are three ($^{13}$N, $^{15}$O and $^{17}$F), and we refer to the sum of the latter as the CNO neutrinos.

Theoretical predictions for the fluxes of these components depend on solar models, which themselves depend on various inputs such as the abundance of heavier ``metal'' elements (specifically all elements heavier than $^4$He, for example $^{12}$C, $^{13}$N and $^{15}$O) i.e. the metallicity. Solar models based on abundances that were inferred from earlier observations and modelling of the photosphere were also in excellent agreement  with helioseismological observations (e.g. GS98 \cite{Basu1997,Basu2004}). However, advances in photosphere and line-formation modelling led to a downward revision of most of the abundances of elements heavier than helium, leading to the more recent ``low-metallicitly'' models~\cite{Asplund:2009fu}, now incompatible with helioseismological data~\cite{2009ASPC..416..193B,Vinyoles:2017bqj}. This disagreement is known as the solar metallicity, or solar abundance problem \cite{Bahcall:2004yr, Bahcall06, Yang07, Basu08, Serenelli:2009yc}, and its solution will require additional, independent data (it may also be a sign of new physics, see e.g. ref.~\cite{Frandsen:2010yj,Vincent:2014jia}). In particular, the CNO neutrino flux is very sensitive to the metallicity of the solar core, as can be seen in Table~\ref{cno_flux_table}, and hence a precise measurement of this flux would help in improving solar models, by providing another metallicity measurement against which they can be tested~\cite{Song:2017kvf,Bergstrom:2016cbh}.

While the components of the proton-proton chain have all been measured (pp~\cite{Bellini:2014uqa}, pep~\cite{Collaboration:2011nga}, $^7$Be~\cite{Bellini:2013lnn}, $^8$B~\cite{Agostini:2017cav,Aharmim:2011vm,Abe:2016nxk}, hep at $1\sigma$~\cite{Bergstrom:2016cbh,Aharmim:2006wq}), a measurement of the CNO neutrino flux has not yet been achieved~\cite{Bergstrom:2016cbh}. This is primarily because CNO neutrinos have neither a high energy, like $^8$B neutrinos, nor a huge flux, like pp neutrinos, but instead form a sub-dominant component of the total solar neutrino spectrum at energies below approximately $1.5$~MeV.
Furthermore, due to effects from the finite energy-resolution and the adopted detection reaction of experiments, the observed spectrum in a detector resulting from CNO neutrinos is expected to resemble strongly that from pep neutrinos, leading to systematic uncertainties in determining the CNO flux even for a background-free experiment.

Many current and future experiments will attempt to measure the CNO flux by looking for neutrinos scattering on electrons, such as Borexino~\cite{Bellini:2013lnn}, SNO+~\cite{Andringa:2015tza}, the Jinping Neutrino Experiment~\cite{JinpingNeutrinoExperimentgroup:2016nol} and liquid argon experiments such as DarkSide and Argo~\cite{Aalseth:2017fik,Franco:2015pha}. It should also be possible to look for CNO neutrinos through scattering with nuclei, in a similar manner to direct searches for dark matter. However this would likely require new technologies to achieve the required low energy threshold~\cite{Strigari:2016ztv}, for example refs.~\cite{Strauss:2017cam,Angloher:2017sxg,Maris:2017xvi,Petricca:2017zdp,Agnese:2017jvy,Schutz:2016tid,Budnik:2017sbu,Aguilar-Arevalo:2016ndq,Arnaud:2017bjh,Agnese:2016cpb,Yang:2017yaw,Arnaud:2017usi}.

In this work we determine the accuracy to which these future experiments can measure the CNO neutrino flux, and whether this is enough to distinguish between the two solar metallicity scenarios.
For experiments looking for electron-recoils from solar neutrinos (e.g. Borexino, Argo and SNO+), some individual projections have been carried out by the experimental collaborations themselves for their respective experiments~\cite{Bellini:2013lnn,Andringa:2015tza,Franco:2015pha}. 
The purpose of this work is not to refute these estimates, but rather to gather the predictions in the same place and to compare as closely as possible the discovery possibilities from a theory perspective. 

Our main aim is to obtain a time-scale for when the CNO flux will be measured. Furthermore, as we will see, it is difficult but not impossible for nuclear recoil experiments to see this flux.  By establishing a time-frame we hope to provide crucial input for experimentalists who plan to run a low-threshold nuclear-recoil experiment, which may also be designed to search for light dark matter.

\begin{table}[bt]
\centering
 \renewcommand{\arraystretch}{1.4}
\begin{tabular}{ p{3cm} || c | c | c  }
\multirow{2}{3cm}{\textbf{Solar Metallicity}} & \multicolumn{3}{c }{\textbf{CNO Neutrino Flux [cm$^{-2}$ s$^{-1}$] }}   \\  
\cline{2-4}
& $^{13}$N [$10^8$] &  $^{15}$O [$10^8$] &  $^{17}$F [$10^6$]  \\  \hline
High & 2.78 $\pm$ 0.42  & 2.05 $\pm$ 0.35  & 5.92 $\pm$ 1.06  \\ \hline
Low & 2.04 $\pm$ 0.29  & 1.44 $\pm$ 0.23 & 3.26 $\pm$ 0.59  \\
\hline
\end{tabular}
\caption{Predicted fluxes of each component of the total neutrino emission from the CNO cycle, used in this work, for the high and low solar metallicity models (GS98 and AGSS09met respectively)~\cite{Vinyoles:2017bqj,Basu1997,Basu2004}.}
\label{cno_flux_table}
\end{table}

\section{Electron-recoil experiments searching for CNO neutrinos}
Experiments searching for electronic recoils induced by interactions with neutrinos will be sensitive to CNO neutrinos, provided their low-energy threshold is below approximately $1.5$~MeV. Here we consider only elastic scattering i.e. $\nu_e + e^- \rightarrow \nu_e + e^-$, and not inelastic scattering as will be seen for example in the DUNE experiment~\cite{Acciarri:2015uup}.

For these experiments there exist various backgrounds whose spectra are similar to that expected from CNO neutrino-induced electronic recoils, and which are often specific to the detector or target. In some cases, the rates of these backgrounds can be determined externally, for example through measurements of the decay rate of daughter nuclei in a radioactive decay chain. It is vital that not only should these background rates be kept as low as possible, to reduce statistical uncertainties, but also ideally that they are known \emph{a priori} to keep systematic uncertainties small.

The aim of our analysis is to quantify the precision with which the CNO flux can be measured for up-coming experimental runs, given the uncertainties on the other solar neutrino fluxes and backgrounds in the energy region of interest. Of particular importance are the systematic uncertainties on determining the CNO neutrino flux, which arise through degeneracies between the CNO flux and both the other neutrino fluxes and the detector-specific background rates. Motivated by this fact, we perform a Markov Chain Monte Carlo (MCMC) parameter scan over all of the neutrino fluxes and background rates, by comparing the spectra of these sources to simulated data with a Poisson likelihood.
Each relevant background rate and neutrino flux has one parameter which determines its total energy-integrated rate, and the spectra are kept fixed up to this normalisation i.e. we fit spectra, but vary only the total rates for backgrounds or fluxes for solar neutrinos.

The MCMC analysis requires that each parameter has with it an associated prior distribution, which reflects the amount of knowledge we have about this parameter before the analysis is performed. For each solar neutrino flux we assume that we have no knowledge and therefore assume a uniform (or flat) prior. In practice all uniform priors are constant between zero and a value much greater than the fiducial value used to generate the simulated data. We will also use a Gaussian prior for some backgrounds when their rates are known to a given precision.

After the analysis is complete the MCMC returns sampled values of each of the parameters, whose histogram is the posterior distribution, which tells us which values of the various parameters fit best to the simulated data. The posterior has a number of dimensions equal to the number of free parameters, and so in order to find the precision with which the CNO flux can be measured we marginalise the posterior over all other parameters.

\begin{figure}[b]
\centering
\includegraphics[width=0.7\textwidth]{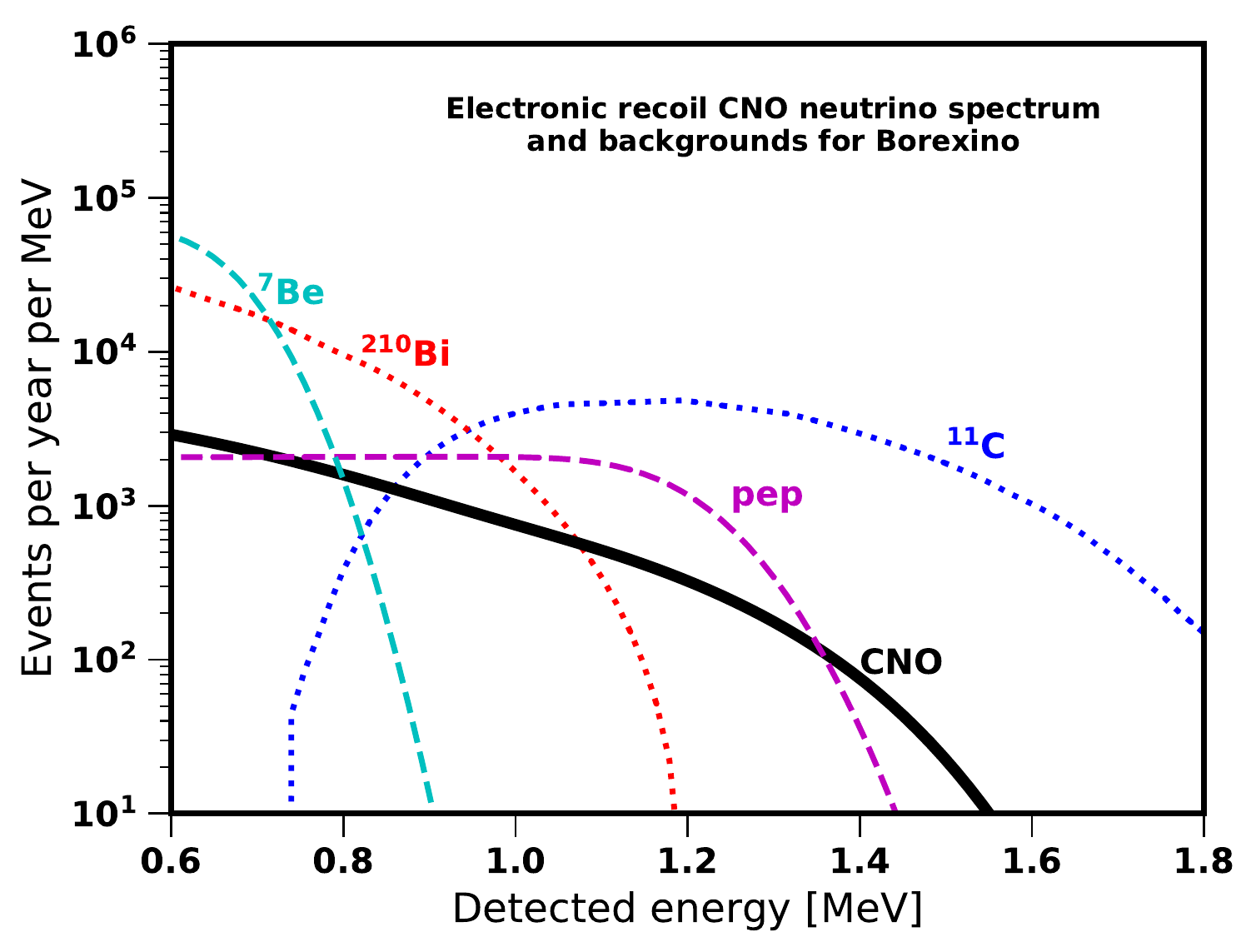}
\caption{Differential electron recoil event rate from CNO neutrinos (solid black) compared with various backgrounds  for Borexino~\cite{Agostini:2017aaa,Collaboration:2011nga,Bellini:2013lnn}. }
\label{fig:spec_borexino}
\end{figure}

\begin{figure*}[t]
\includegraphics[width=0.99\textwidth]{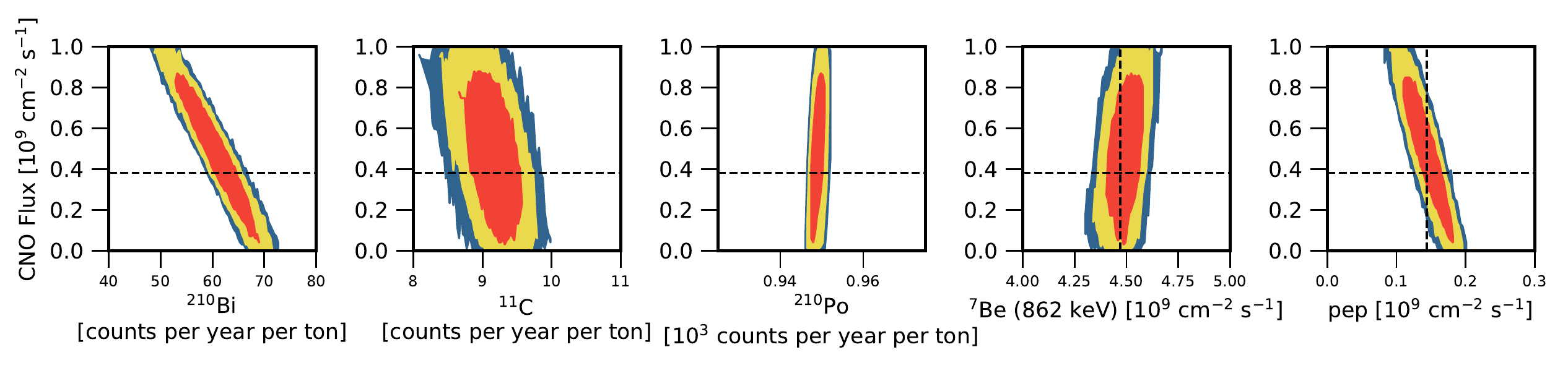} \\
\includegraphics[width=0.99\textwidth]{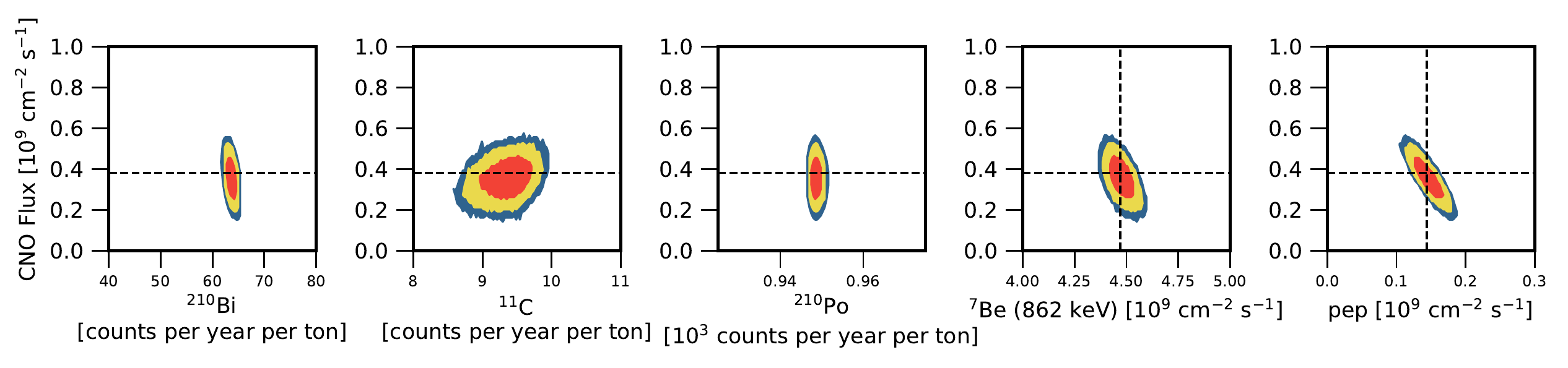}
\caption{Two-dimensional marginalised posterior distributions for Borexino running for 10~years with the pessimistic prior set, assuming $10\%$ uncertainty on the $^{210}$Bi rate (top) or the optimistic prior set assuming $1\%$ uncertainty on the $^{210}$Bi rate  (bottom).The dashed black lines show the fiducial values used to generate simulated data. In each panel the CNO flux is compared to different background rates and other solar neutrino fluxes, where the red  contour bounds $68\%$ of the posterior distribution, while yellow bounds $95\%$ and blue $99\%$.}
\label{fig:borexino_mcmc}
\end{figure*}

In the rest of this section we detail, for each experiment, the detector-specific backgrounds and their priors, and the results of our MCMC analysis on the precision with which the CNO neutrino flux can be measured. We consider Borexino, SNO+ and dual-phase liquid argon experiments such as DarkSide or Argo. We do not perform a dedicated analysis for the Jinping Neutrino Experiment. However its projected sensitivity should be similar to that of SNO+, corrected for the different target mass~\cite{JinpingNeutrinoExperimentgroup:2016nol}.

\subsection{Borexino}

Borexino is a liquid scintillator experiment situated in the Gran Sasso laboratory, with a target mass of 278 tons and a fiducial mass in the most recent run of 71.3 tons~\cite{Bellini:2013lnn,Agostini:2017ixy}. Neutrinos are detected by observing scintillation light from the electrons, which have been given MeV-levels of energy though elastic scattering.
In their previous Phase-I run, the Borexino collaboration did not observe CNO neutrinos and so set a limit on the CNO interaction rate of $R_{\mathrm{CNO}} < 7.9~[100 \, \mathrm{ton} \, \mathrm{day}]^{-1}$ at $95\%$ confidence~\cite{Bellini:2013lnn}, with a similar limit set in Phase-II~\cite{Agostini:2017ixy}. Here we consider a future run of Borexino, taking into account various improvements as detailed below.
As the only running experiment considered in this work, it will have the most realistic projections for measuring the CNO flux, since for example its backgrounds and energy-resolution are well-understood already.

As shown in figure~\ref{fig:spec_borexino}, apart from the other solar neutrino fluxes ($^7$Be and pep) the largest backgrounds in the region of interest for a CNO neutrino search by Borexino arise from $^{210}$Bi beta-decays (originating from the slow decay of $^{210}$Pb) and $^{11}$C decay to positrons~\cite{Agostini:2017ixy,Agostini:2017aaa,Collaboration:2011nga}. The former is particularly troublesome as its spectrum follows closely that expected from the CNO neutrinos, and so any measurement of the CNO neutrino rate will be partially degenerate with the rate of $^{210}$Bi beta-decay electrons. 
For Borexino, it has been suggested that the magnitude of this $^{210}$Bi background can be inferred to an accuracy of $10\%$ (i.e. around $6.4$ counts per ton per year) or better through measurements of the alpha-decay of its daughter nucleus $^{210}$Po~\cite{Villante:2011zh}. Recent upgrades to the Borexino experiment, to establish secular equilibrium in the $^{210}$Pb decay chain, mean that such a measurement is now possible~\cite{calaprice_talk_borexino}.
The $^{11}$C positron background arises from cosmogenic muons interacting with $^{12}$C nuclei. Its contamination in the region of interest for a CNO search is reduced by an order of magnitude using ``three-fold'' coincidence cuts on the cosmogenic muons and neutrons which are commonly produced with the $^{11}$C nuclei~\cite{Agostini:2017ixy,Agostini:2017aaa,Collaboration:2011nga}.

\begin{table}[tb]
\centering
 \renewcommand{\arraystretch}{1.2}
\begin{tabular}{ p{2.5cm} || c | c | c | c }
\multirow{3}{3cm}{\textbf{Background}} & \multicolumn{2}{c |}{\textbf{Optimistic}} & \multicolumn{2}{c}{\textbf{Pessimistic}}  \\  
\cline{2-5}
& Value & Error & Value & Error  \\ 
& [(ton yr)$^{-1}$] & [1$\sigma$] & [(ton yr)$^{-1}$] & [1$\sigma$] \\ \hline
$^{210}$Bi& 64 & $1\%$ & 64 & $10\%$  \\
\hline
$^{210}$Po& 950 & Free & 950 & Free \\ \hline
$^{11}$C& 9.49 & Free & 9.49  & Free \\ 
\hline
\end{tabular}
\caption{Fiducial values and relative uncertainties for the backgrounds in Borexino based on Phase-II data~\cite{Agostini:2017ixy}. We provide either the one-sigma error on the 
Gaussian prior, with central value equal to the fiducial value, or allow ``Free'' errors, meaning the prior distribution is uniform and unconstrained.}
\label{borexino_priors}
\end{table}

In order to understand the projected sensitivity of Borexino to CNO neutrinos in the future, and the effect of this technique to measure the $^{210}$Bi background, we perform two MCMC runs each with a different prior distribution for the $^{210}$Bi rate. In the first case, we assume that the method proposed in ref.~\cite{Villante:2011zh} allows the $^{210}$Bi background rate to be measured to a precision of $10\%$ of the total rate (i.e. around $6.4$ counts per ton per year), and so the $^{210}$Bi rate is given a Gaussian prior with this value as its one-sigma standard deviation. While in the second case we assume an optimistic scenario in which the $^{210}$Bi rate will be measured to $1\%$ accuracy (i.e. around $0.64$ counts per ton per year).  Our different priors and background rates are summarized in Table~\ref{borexino_priors}, based on the Phase-II\footnote{We use Phase-II as this is the most recent published Borexino data, though note that the backgrounds in the next phase of Borexino will likely be smaller due to the slow decay of $^{210}$Pb and its daughters.} run in ref.~\cite{Agostini:2017ixy}, which we also use to obtain the spectra of the background components.

Figure~\ref{fig:borexino_mcmc} shows a comparison of the posterior distributions from our MCMC projection for Borexino with either the optimistic or pessimistic prior sets, assuming $10$~years of data-taking. Each panel is a two-dimensional projection of the total posterior distribution i.e. the posterior distribution summed over all free parameters except those on the $x$ and $y$ axes. This allows us to see degeneracies between the CNO flux and the different backgrounds and other solar neutrino spectra. For example, the left-most panel on each plot shows that the CNO flux measurement is strongly degenerate with the $^{210}$Bi background, as expected since they have similar spectra. The precision on the CNO flux measurement is then obtained by marginalising over the other parameters, for example by summer over the $x$-axis in any of the two-dimensional panels, and so projecting onto the $y$-axis.

For the pessimistic priors there is a wide range of pairs of values for the $^{210}$Bi rate and CNO flux which provide a good fit to the data, which means that we have only weak constraints on the CNO flux, since it can be compensated by changing the $^{210}$Bi rate. However, with the optimistic prior set values of the $^{210}$Bi rate close to its fiducial value are strongly preferred, since we know those values further away can not be physical as we assume the $^{210}$Bi rate has been measured to $1\%$ accuracy. Hence the degeneracy is broken and the allowed range of CNO flux values is much smaller.  Importantly for Borexino this degeneracy is very strong, which is why the effect of the optimistic priors is so clear, indicating that the ability of Borexino to measure the CNO flux will be extremely sensitive to the precision with which the $^{210}$Bi rate is measured, as well as to the total rate of this background.

Similarly, in the right-most panel we show the degeneracy between the CNO and pep neutrinos fluxes, resulting from their similar elastic-scattering spectra. This significantly degrades the ability of Borexino, and indeed all neutrino detectors, to measure the CNO flux. For example, if we knew the pep neutrino flux precisely then the CNO flux could be constrained to be within the region where the vertical dashed line in the right-most panel intersects with the shaded regions, projected onto the $y$-axis, but unfortunately disentangling these two neutrinos fluxes is not possible. 

\subsection{SNO+}
\begin{figure}[b]
\centering
\includegraphics[width=0.7\textwidth]{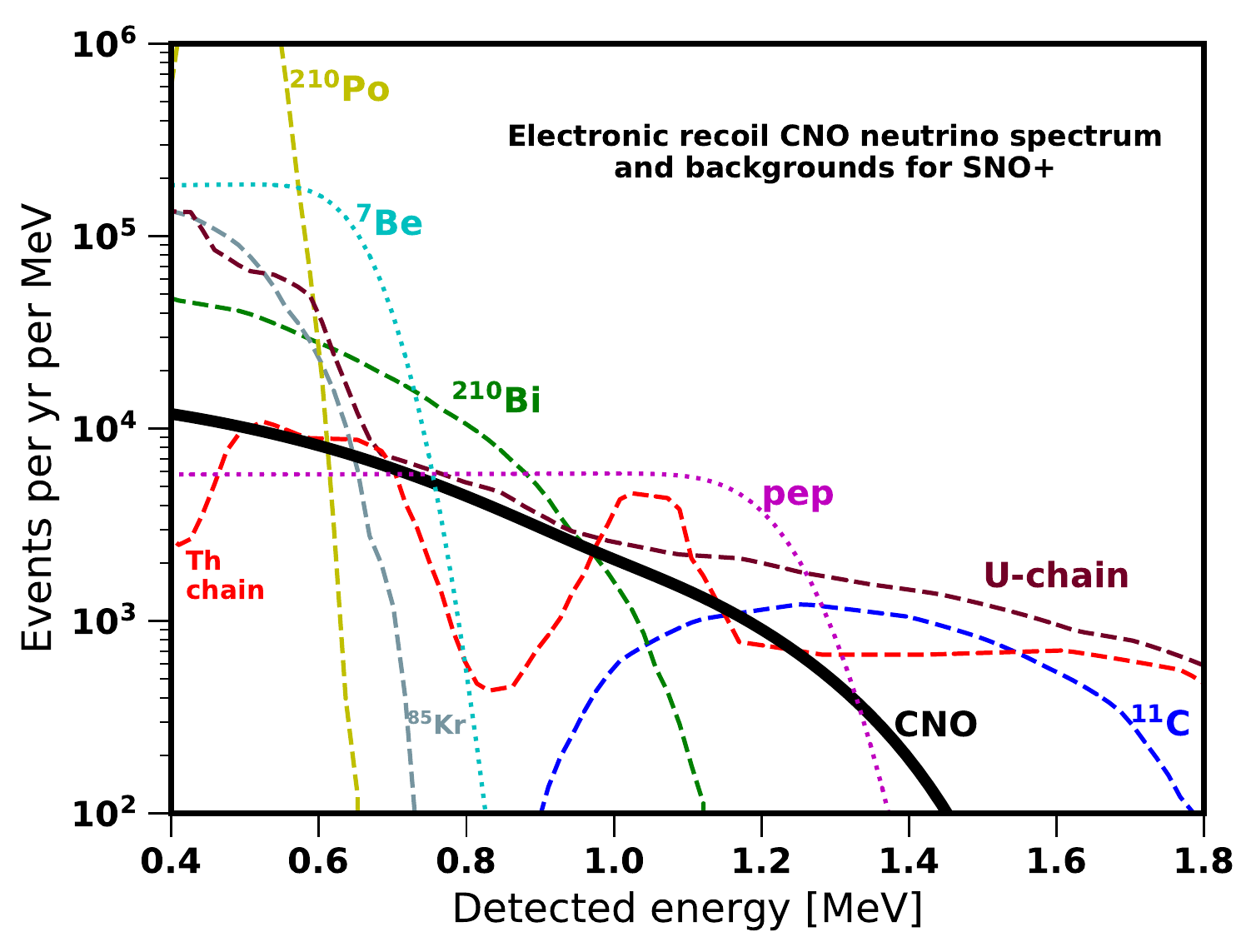}
\caption{The spectrum of electronic recoil events from CNO neutrinos (solid black) compared with various backgrounds expected for SNO+~\cite{Andringa:2015tza}. }
\label{fig:spec_snoplus}
\end{figure}

\begin{figure*}[t]
\includegraphics[width=0.99\textwidth]{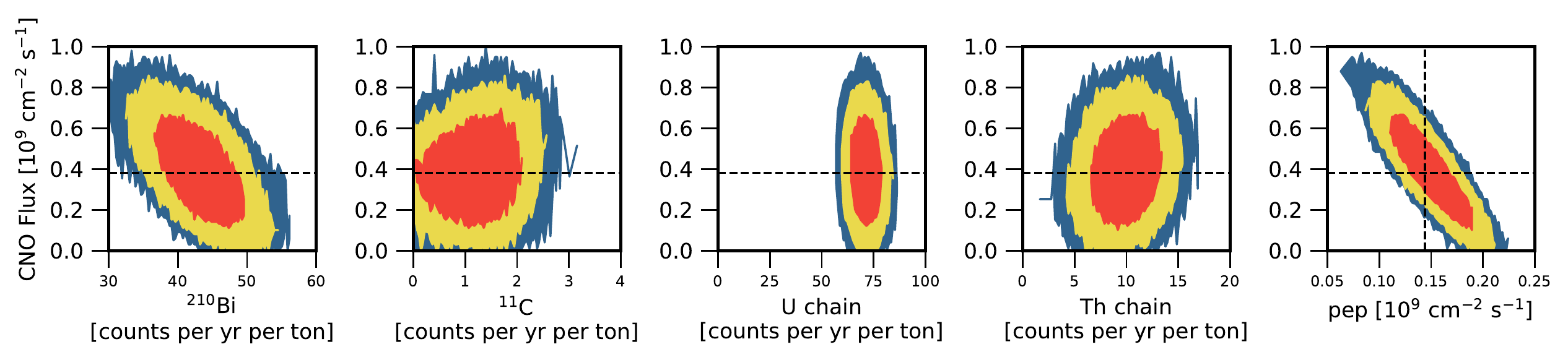} \\
\includegraphics[width=0.99\textwidth]{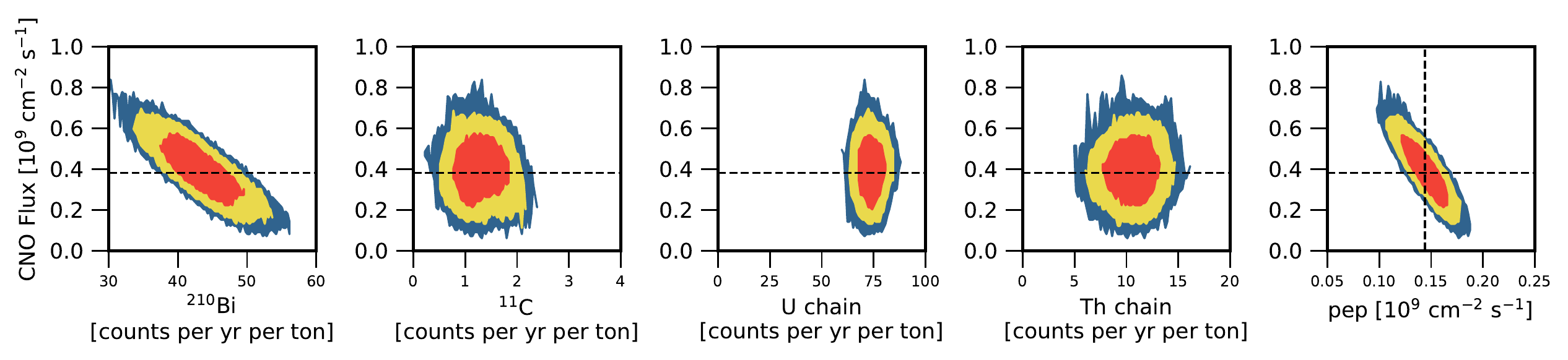}
\caption{Marginalised posteriors for SNO+ with 6 months (top) or 3 years (bottom) of data and pessimistic priors. A diagonal shaped distribution means the parameters have some degeneracy between them. The dashed black lines show the fiducial values used to generate simulated data. In each panel the CNO flux is compared to different background rates and other solar neutrino fluxes, where the red  contour bounds $68\%$ of the posterior distribution, while yellow bounds $95\%$ and blue $99\%$. Note that all total rates are calculated above an energy of $0.3$~MeV.}
\label{fig:sno_plus}
\end{figure*}

SNO+ is an upcoming liquid scintillator experiment situated at SNOLAB, with a 780 ton target mass~\cite{Andringa:2015tza}. 
The expected neutrino signals and backgrounds are shown in figure~\ref{fig:spec_snoplus}, where following ref.~\cite{Andringa:2015tza} the background rates are based on those already achieved in Borexino. Due to its location within a deeper site compared with Borexino, its cosmogenic backgrounds such as $^{11}$C are smaller. The primary purpose of the SNO+ experiment is to measure neutrino-less double-beta decay using $^{130}$Te loaded into the liquid scintillator~\cite{Andringa:2015tza}. Due to this, the SNO+ detector may only be sensitive to CNO neutrinos for a fraction of its total running time, since this $^{130}$Te beta-decay leads to a large background in the CNO neutrino energy range. Hence in this work we consider only SNO+ in its pure scintillator mode i.e. without $^{130}$Te doping.

As with Borexino, and shown in figure~\ref{fig:spec_snoplus}, the SNO+ detector will have a difficult background from $^{210}$Bi, although it should have better energy-resolution making this background easier to separate from the CNO neutrino spectrum~\cite{Andringa:2015tza}. It is also expected to have backgrounds originating from the thorium and uranium decay chains and $^{85}$Kr, $^{210}$Po and $^{11}$C. As detailed in ref.~\cite{Andringa:2015tza}, each of these backgrounds can be measured to some extent using coincidence decays and tagging of daughter nuclei. For $^{210}$Po, $\alpha$-tagging will be used to reduce this background by $95\%$. Hence we adopt Gaussian priors on these background rates which reflect the expected precision to which these backgrounds will be determined, as shown in table~\ref{sno_plus_priors}. We perform two runs, an optimistic and pessimistic scenario, where in the former case the  $^{210}$Bi background can be measured to an accuracy of $1\%$, while in the latter case it will be measured to $10\%$ precision, both employing the method proposed in ref.~\cite{Villante:2011zh}.
It remains to be seen
whether the SNO+ background projections
are actually achievable in practice.

\begin{table}[t]
\centering
 \renewcommand{\arraystretch}{1.2}
\begin{tabular}{ p{3cm} || c | c | c | c }
\multirow{3}{3cm}{\textbf{Background}} & \multicolumn{2}{c |}{\textbf{Optimistic}} & \multicolumn{2}{c}{\textbf{Pessimistic}}  \\  
\cline{2-5}
& Value & Error & Value & Error  \\ 
& [(ton yr)$^{-1}$] &[1$\sigma$]  & [(ton yr)$^{-1}$] & [1$\sigma$] \\ \hline
$^{210}$Bi& 45.4 & $1\%$ & 45.4 & $10\%$  \\
\hline
$^{210}$Po& 1530 & $20\%$ & 1530 & $20\%$ \\ \hline
$^{11}$C& 1.74 & Free & 1.74  & Free \\ \hline
$^{85}$Kr& 96.4 & $50\%$ & 96.4  & $50\%$ \\ \hline
U chain& 74.4 & $7\%$ & 74.4  & $7\%$ \\ \hline
Th chain& 11.1 & $25\%$ & 11.1  & $25\%$ \\ 
\hline
\end{tabular}
\caption{Fiducial values and relative uncertainties for the backgrounds in SNO+ above a threshold energy of $0.3$~MeV (without $^{130}$Te doping)~\cite{Andringa:2015tza}. A percentage uncertainty means the one-sigma error on the 
Gaussian prior, with central value equal to the fiducial value. Columns with ``Free'' errors mean the prior distribution is uniform
i.e. unconstrained.}
\label{sno_plus_priors}
\end{table}

Figure~\ref{fig:sno_plus} shows the results of our MCMC analysis for SNO+, assuming a $50\%$ fiducial mass cut. There is a significant systematic uncertainty on the CNO flux arising from the $^{210}$Bi background, as can be seen by the degeneracy between these two parameters in the left-most panel. The uncertainty on the CNO flux alone is then obtained by marginalising over the $^{210}$Bi rate, and all of the other parameters. We have made the assumption that the $^{210}$Bi background rate can be measured to an accuracy of $10\%$ i.e. the pessimistic priors, which makes the CNO-$^{210}$Bi degeneracy clear. With six months of data, the degeneracy is such that a zero CNO flux can not be excluded with high significance.

The improvement on the CNO flux measurement going from 6 months to 3 years of SNO+ data is large. The reason for this is subtle, as more data also helps relieve the partial degeneracy between both $^{210}$Bi and pep and CNO, especially when the tails of their spectra can be measured precisely. This can be seen in the left-most panel of each plot, where the best-fit region between CNO and $^{210}$Bi rotates towards the horizontal, meaning the degeneracy between these two parameters has been reduced. This is because the increased statistics means that the spectra of the CNO and $^{210}$Bi components, as shown in figure~\ref{fig:spec_snoplus}, can be more easily distinguished. Hence an increased amount of data has not only reduced statistical uncertainties, but also systematics .
There is little degeneracy between the CNO flux and the $^{11}$C or U-chain backgrounds, since their spectra are not expected to be similar to that from CNO. though there is a small amount of degeneracy for the 3~year run between the Th-chain and CNO.

\subsection{Liquid argon TPCs}
\begin{figure}[b]
\centering
\includegraphics[width=0.7\textwidth]{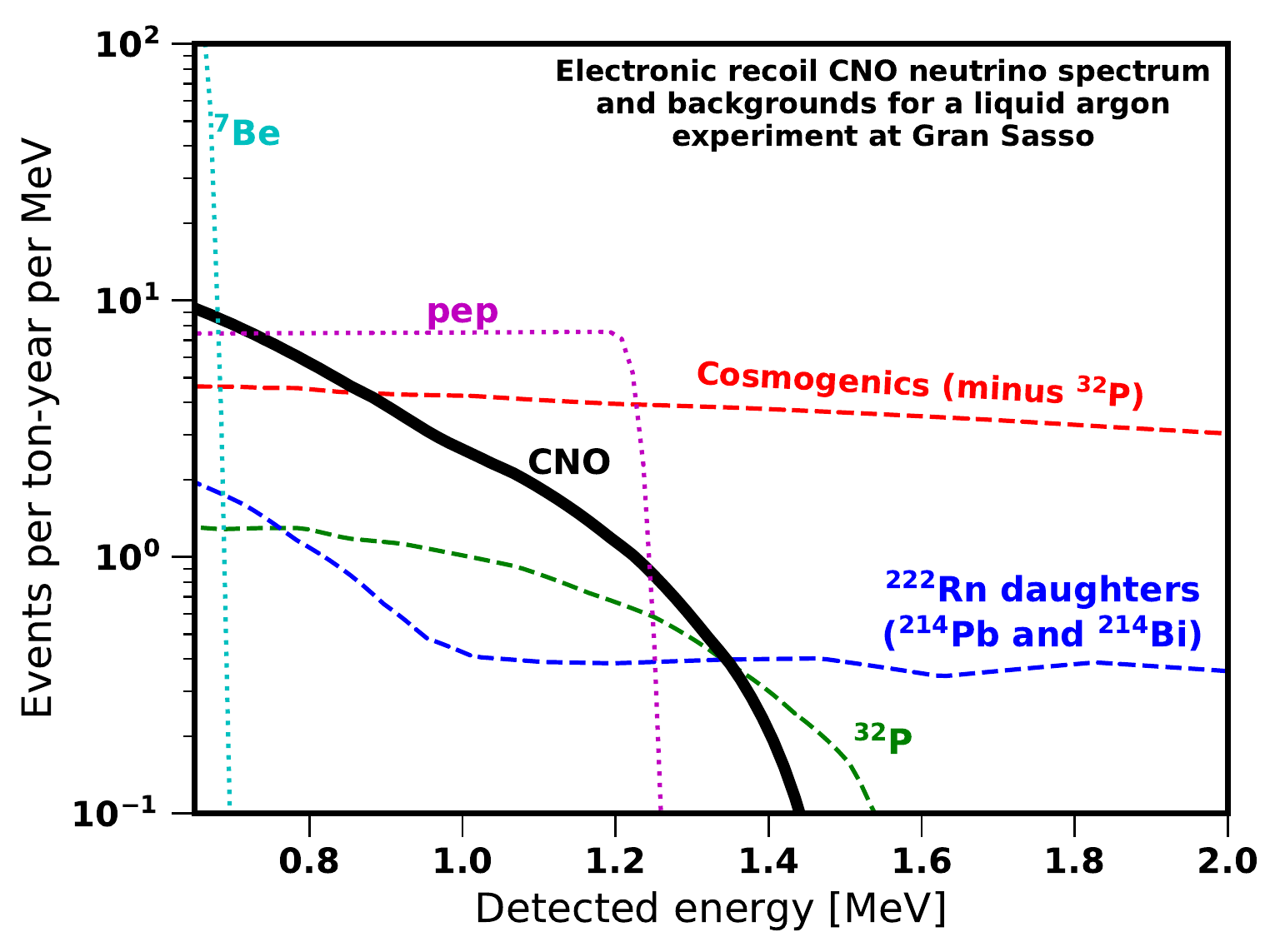}
\caption{The spectrum of electronic recoil events from CNO neutrinos (solid black) compared with various backgrounds expected for a liquid argon experiment operating in the Gran Sasso lab~\cite{Franco:2015pha}. The $^{222}$Rn contamination is assumed to be $10$~$\mu$Bq per $100$~ton. }
\label{fig:spec_argon_gs}
\end{figure}

\begin{figure*}[t]
\includegraphics[width=0.99\textwidth]{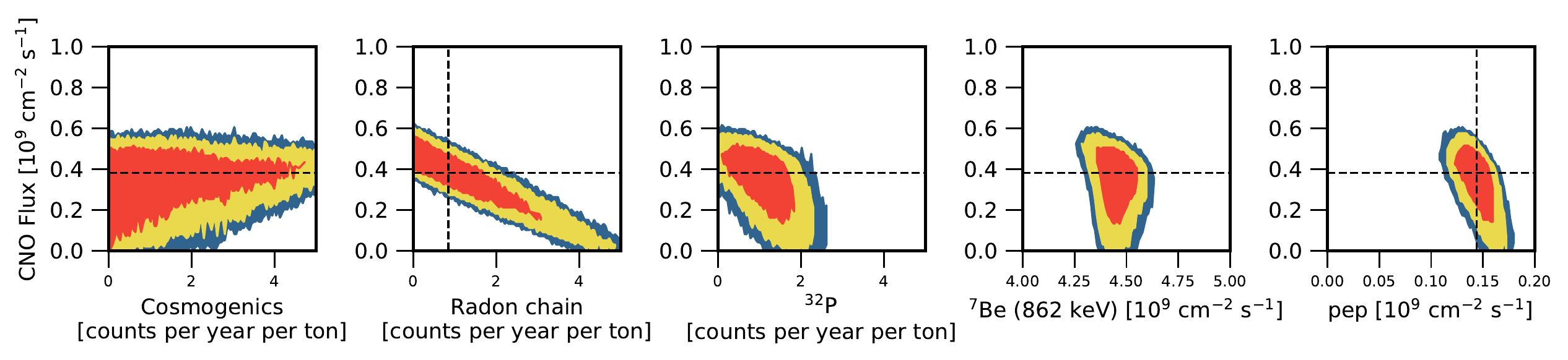} \\
\includegraphics[width=0.99\textwidth]{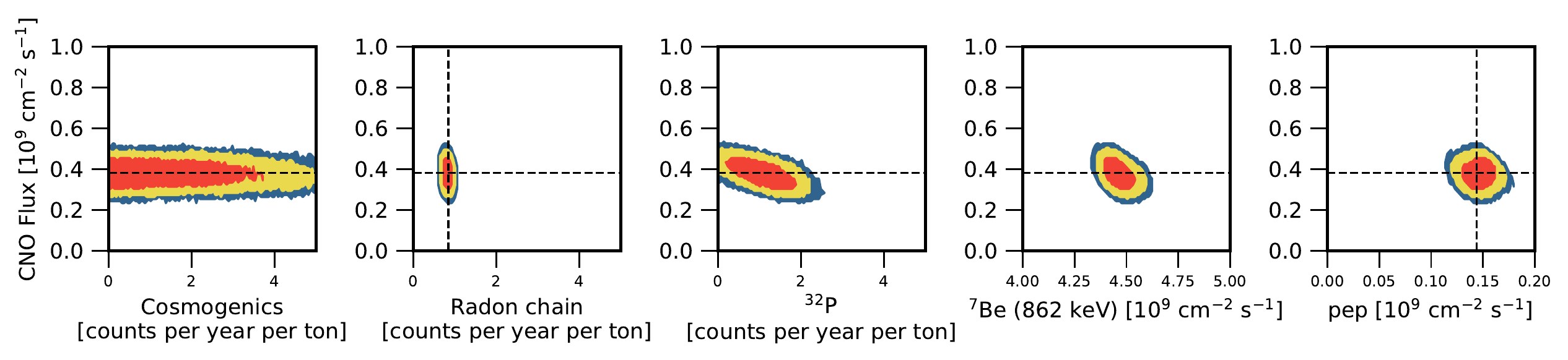} \\
\caption{Contours showing the degeneracy between the CNO flux and various backgrounds and other solar neutrino fluxes for a liquid argon electronic-recoil experiment running in the Gran Sasso lab, with an exposure of 1000 ton-years with an unknown radon background (top) or a radon background known to a precision of $10\%$ at one-sigma (bottom). The dashed black lines show the fiducial values used to generate simulated data. In each panel the red  contour bounds $68\%$ of the posterior distribution, while yellow bounds $95\%$ and blue $99\%$.}
\label{fig:argon_GS_1000tonyr}
\end{figure*}

In addition to the currently-running DEAP-3600~\cite{Amaudruz:2017ekt} there exist several upcoming or proposed experiments, such as DarkSide-20k (with a fiducial mass of 20~tons) and Argo (with a fiducial mass of 100~tons)~\cite{Aalseth:2017fik,Franco:2015pha}, based on a dual-phase argon\footnote{We do not consider liquid xenon experiments for an electronic recoil CNO neutrino search due to their large backgrounds, especially from $^{136}$Xe double-beta decay~\cite{Baudis:2013qla}.} time-projection chamber (TPC) set-up which will look for both dark matter, primarily through nuclear-recoils, and neutrinos~\cite{Aalseth:2017fik}. Here we take a more general view and will not look at a specific experiment in detail, with the motivation of determining the best experimental set-up to maximize sensitivity to CNO neutrinos for an argon experiment. Additionally, although we know that DarkSide-20k will be housed in the Gran Sasso lab~\cite{Aalseth:2017fik}, we do not know if this will be the case for Argo, and so we will consider also the case where Argo is housed in SNOLAB or Jinping~\cite{JinpingNeutrinoExperimentgroup:2016nol}. 

The analysis of ref.~\cite{Franco:2015pha} identified two major sources of backgrounds relevant to a search for CNO neutrinos using electronic recoils in a liquid argon experiment: cosmogenic backgrounds and $^{222}$Rn  from the detector environment. The background from $^{222}$Rn comes from $\beta$-decay electrons emitted by $^{214}$Pb and $^{214}$Bi, two of the daughters of $^{222}$Rn $\alpha$-decay. In this work we use the background spectra calculated in ref.~\cite{Franco:2015pha}, which we show in figure~\ref{fig:spec_argon_gs}. 
As can be seen from
the figure, the excellent energy resolution
makes the pep spectrum easy to distinguish
from CNO.

The size of the background from the $^{222}$Rn decay chain depends on the specific detector set-up of the argon experiment, and needs to be smaller than $\sim 100$~$\mu$Bq per $100$~ton to make a measurement of the CNO flux achievable~\cite{Franco:2015pha}. The DEAP-3600 experiment had a $^{222}$Rn contamination level of $(1.8 \pm 0.2) \cdot 10^4$~$\mu$Bq per $100$~ton in its most recent run~\cite{Amaudruz:2017ekt}, however a larger experiment such as DarkSide-20k or Argo should have a smaller contamination by volume, as the $^{222}$Rn enters through the walls of the argon tank and scales less strongly with increasing experiment size than the total target mass. Throughout this work we assume a $^{222}$Rn contamination of $10$~$\mu$Bq per $100$~ton.
The magnitude of the radon background can be measured through observations of the delayed coincidence between $^{214}$Bi $\beta$-decay and $^{214}$Po $\alpha$-decay, another daughter of $^{222}$Rn $\alpha$-decay~\cite{Franco:2015pha,Amaudruz:2017ekt}. Hence in order to understand the effect of this measurement on the CNO sensitivity, we perform analyses with the $^{222}$Rn freely-varying and with it fixed \emph{a priori} to a given precision. All of the fiducial values and priors we use in our analysis are given in table~\ref{argon_priors}.

\begin{table}[b]
\centering
 \renewcommand{\arraystretch}{1.2}
\begin{tabular}{ p{3.5cm} || c | c | c | c }
\multirow{3}{3.5cm}{\textbf{Background}} & \multicolumn{2}{c |}{\textbf{Optimistic}} & \multicolumn{2}{c}{\textbf{Pessimistic}}  \\  
\cline{2-5}
& Value & Error & Value & Error  \\ 
& [(ton yr)$^{-1}$] &  & [(ton yr)$^{-1}$] &  \\ \hline
$^{222}$Rn daughters& 0.84 & $10\%$ & 0.84 & Free  \\
\hline
$^{32}$P& 0.78  & Free & 0.78 & Free \\ \hline
Other cosmogenics& 5.4 & Free & 5.4  & Free \\ 
\hline
\end{tabular}
\caption{Fiducial values and uncertainties for the backgrounds in a liquid argon experiment, based in Gran Sasso~\cite{Franco:2015pha}, where ``Other cosmogenics" refers to all cosmogenic backgrounds except for $^{32}$P. A percentage uncertainty means the one-sigma error on the 
Gaussian prior, with central value equal to the fiducial value. Columns with ``Free'' errors mean the prior distribution is uniform
i.e. unconstrained.}
\label{argon_priors}
\end{table}

\begin{figure*}[t]
\includegraphics[width=0.99\textwidth]{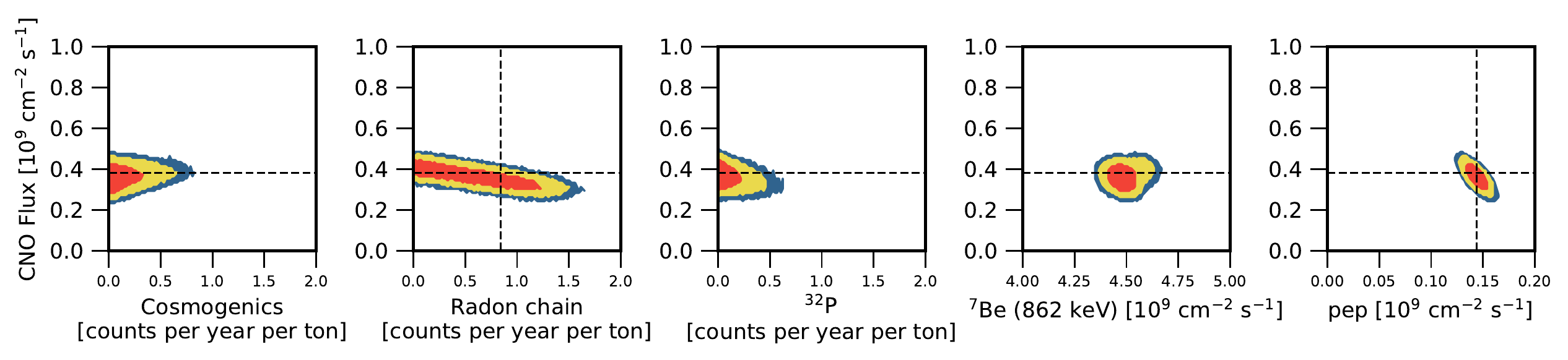} \\
\includegraphics[width=0.99\textwidth]{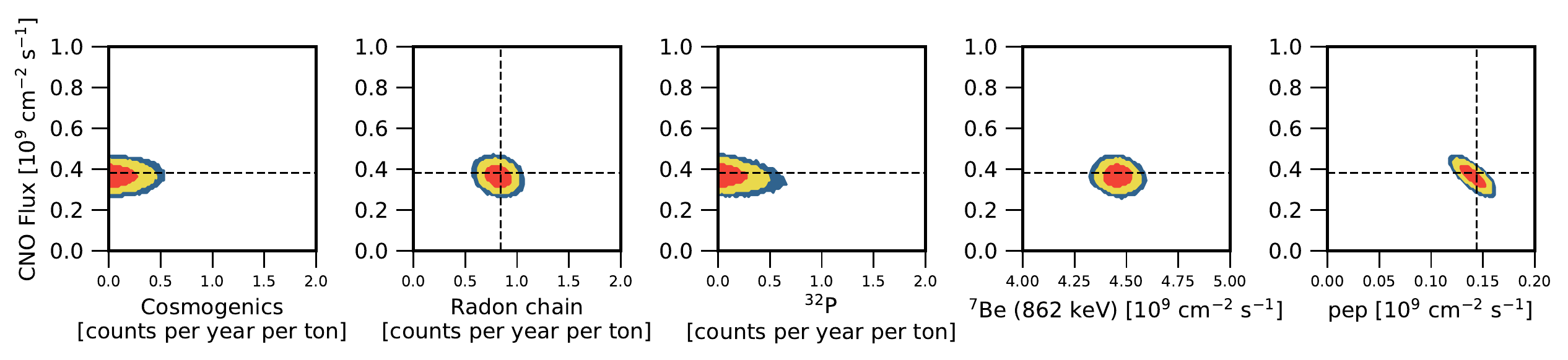} 
\caption{Contours showing the degeneracy between the CNO flux and various backgrounds and other solar neutrino fluxes for a liquid argon electronic-recoil experiment running in SNOLAB or Jinping, with an exposure of 1000 ton-years with an unknown radon background (top) or a radon background known to a precision of $10\%$ at one-sigma (bottom). The dashed black lines show the fiducial values used to generate simulated data. In each panel the red  contour bounds $68\%$ of the posterior distribution, while yellow bounds $95\%$ and blue $99\%$.}
\label{fig:argon_SNOlab_1000tonyr}
\end{figure*}

The result of our MCMC scan for a liquid argon experiment based in Gran Sasso lab with a $1000$~ton-year exposure is shown in figure~\ref{fig:argon_GS_1000tonyr}. It is clear that the CNO flux is degenerate with measurements of the radon background and the background from $^{32}$P decays, as expected from their similar spectra at lower energies seen in figure~\ref{fig:spec_argon_gs}. Indeed, as can be seen in the second panel from the left in each plot, knowledge of the radon background vastly improves the precision to which the CNO flux can be measured, by eliminating regions of parameter space where a larger radon background can compensate for a smaller CNO flux, and vice versa. For the pessimistic case, the $99\%$ region extends to values where the CNO flux is zero, and the radon background is around $5$ counts per day per ton, and so when marginalising over the radon background SNO+ can not exclude the possibility of a zero CNO flux at $99\%$ confidence. By contrast, when we fix the radon background for the optimistic case, a zero CNO flux will be strongly disfavoured by the experiment. Hence in order to measure the CNO neutrino flux at high-precision with an argon-based experiment, it is crucial that the radon background is kept as small as possible and is measured.

In the case where the liquid argon experiment, such as Argo, is based in SNOLAB or Jinping, we assume that the $^{32}$P and ``Other cosmogenics'' background rates in Table~\ref{argon_priors} are 100 times lower owing to the smaller atmospheric muon flux~\cite{JinpingNeutrinoExperimentgroup:2016nol}. As shown in figure~\ref{fig:argon_SNOlab_1000tonyr}, the improvement gained by this reduction in the cosmogenic background is larger than one might expect, and is due to two effects: the first is the reduction in statistical uncertainty from the lower total background. The second is a partial breaking of the degeneracy between the CNO and $^{222}$Rn spectra, as the radon background spectrum has a predominant tail which will be difficult to measure above the cosmogenic background in Gran Sasso (see figure~\ref{fig:spec_argon_gs}), but is easily visible for Argo, or a similar experiment, based in SNOLAB or Jinping. This also means that an external measurement of the $^{222}$Rn background through coincidence decays is less important for an experiment based in SNOLAB or Jinping, since its rate can be determined well from the beta-spectrum alone. Indeed as can be seen in the second-from-left panel in figure~\ref{fig:argon_SNOlab_1000tonyr} there is still degeneracy between the CNO flux and the $^{222}$Rn rate in the pessimistic case, but it no longer extends to values of zero for the CNO flux. This is due to the fact that a zero CNO flux  would need to be compensated by a larger radon background which would also introduce too many events at higher energies above the cut-off of the CNO spectrum, while in the Gran Sasso case these extra events would not be easily visible above the large cosmogenic background.

\subsection{Comparison of potential CNO flux measurements}

\begin{figure*}[t]
\includegraphics[width=0.98\textwidth]{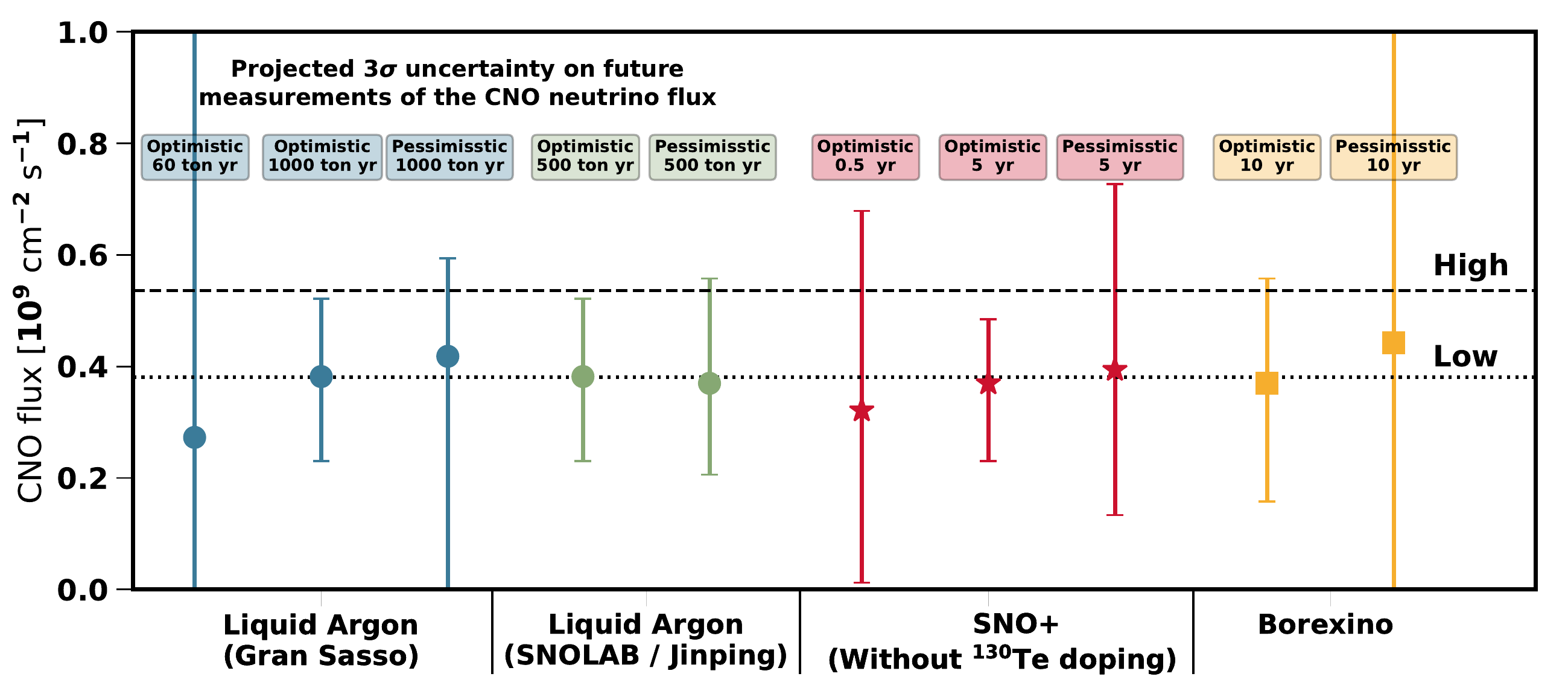}
\caption{Comparison of experiments searching for CNO neutrinos using electronic recoils, with either exposure or running-time labelled. The error-bars show the projected $3 \sigma$ precision with which each experimental run will be able to measure the flux of CNO neutrinos.
The optimistic and pessimistic scenarios differ by how accurately key backgrounds will be measured, and are detailed in Table~\ref{argon_priors} for liquid argon experiments (e.g. DarkSide-20k or Argo)~\cite{Aalseth:2017fik,Franco:2015pha}, Table~\ref{sno_plus_priors} for SNO+~\cite{Andringa:2015tza} and Table~\ref{borexino_priors} for Borexino~\cite{Bellini:2013lnn,Agostini:2017ixy}. The horizontal lines labelled ``High" and ``Low" refer to the high and low metallicity scenarios respectively~\cite{Vinyoles:2017bqj,Basu1997,Basu2004}. Each column and colour represents a different experiment or lab combination.}
\label{fig:ER_comparison}
\end{figure*}

Figure~\ref{fig:ER_comparison} shows the expected precision with which our various projected experimental runs can measure the CNO neutrino flux at $3 \sigma$ confidence. For each run, the error bars bracket the region where the value of the CNO neutrino flux provides a fit to the simulated data which deviates by $3 \sigma$ or less from the best-fit value, marginalising over the other parameters in the MCMC fit. We have chosen $3 \sigma$ since it is generally considered to be a sufficient level of confidence for discovery in astrophysics. 

A liquid argon electronic recoil experiment based in Gran Sasso lab will obtain the required precision with a $1000$~ton~year exposure and a radon background which has been measured \emph{a priori} to a precision of $10\%$ at $1 \sigma$ (or better). This exposure could be obtained by the proposed Argo experiment in just over $3$~years if it has a mass of $300$~tons, but will likely take up to $10$~years since the plan is to use $100$~tons as the fiducial mass~\cite{Davini:2016vpd,Aalseth:2017fik,Franco:2015pha}. With such an exposure Argo in Gran Sasso will be able to distinguish between the low and high metallicity scenarios for the CNO flux at $3 \sigma$ confidence, though as expected from figure~\ref{fig:argon_GS_1000tonyr} this is not the case if the size of the radon background is unknown or poorly measured. Unfortunately for a lower exposure the precision degrades significantly, and so there is no prospect for measuring the CNO flux with DarkSide-20k. 
In all such instances the argon experiments
gain a significant advantage from their excellent energy resolution.

For Argo, or a similar experiment, based in SNOLAB or Jinping the prospects are even better. This is mainly because the $^{222}$Rn background is much easier to distinguish from the CNO spectrum, compared with Gran Sasso where the larger cosmogenic background makes it difficult to precisely measure the radon from its beta-spectrum alone. In this case a precise determination of the CNO flux can be made in approximately half the time, with a $500$~ton~year exposure, even if the radon background is not measured using observations of coincidence decays. This would take Argo (in SNOLAB or Jinping) $5$~years with a $100$~ton fiducial mass. In both cases this requires a radon background at the level of $10 \mu$Bq per 100 tons contamination.

SNO+ in its pure liquid-scintillator mode (i.e. without $^{130}$Te added) should achieve a good measurement of the CNO flux with three years of running, and with five years of data will be able to distinguish the two scenarios for the solar metallicity using CNO neutrinos at $3 \sigma$, provided that the $^{210}$Bi background can be measured to an accuracy of $1\%$. As can be seen in figure~\ref{fig:ER_comparison}, without this constraint on the $^{210}$Bi background rate, the measurement of the CNO flux will be much less precise due to the large degeneracy between the CNO flux and the $^{210}$Bi rate. Hence an accurate determination of the $^{210}$Bi rate in SNO+ is crucial for a CNO neutrino search.
A SNO+ run of six months should be able to detect CNO neutrinos with 99\% confidence, but will lead to only a modest constraint on the absolute flux, which is unlikely to be better than the limit already set by Borexino~\cite{Bellini:2013lnn,Agostini:2017ixy}. Hence SNO+ would need to run for between $3$ and $5$ years to be confident of measuring the CNO flux to enough precision to solve the solar metallicity problem, provided that its background levels are as low as those already measured in Borexino.

After ten years of running, Borexino could measure the CNO flux with enough precision to separate the two solar models, provided that the $^{210}$Bi background is measured to a precision of $1\%$ i.e. the optimistic prior scenario.
However, this projection is extremely sensitive to the level of precision to which the backgrounds, especially $^{210}$Bi can be measured to i.e. for Borexino the difference in CNO flux precision between the optimistic and pessimistic cases is particularly large.
If the future run of Borexino has larger background rates than we have assumed or if these are not measured to the precision assumed in our optimistic prior case, then Borexino is unlikely to measure the CNO flux even after ten years, as is clear from the large error bars for the pessimistic case.

Note also that we have only fit energy spectra in our analysis of each experiment, while the experimental collaborations will have access to additional information. Hence our projections should be considered as conservative estimates. 
For example, the Borexino collaboration will have access to more data such as the spatial position of each interaction, information on coincidences between different detected events and the pulse-shape of each event, which may improve their sensitivity to CNO neutrinos~\cite{Agostini:2017ixy,Agostini:2017aaa,Collaboration:2011nga}. The SNO+ and Argo groups would likely have access to similar information, though in all cases the improvement is not likely to be large, considering that we have made efforts to implement the effects of these cuts where possible.

Which of these experimental runs gets to the CNO flux measurement first, especially with enough precision to solve the solar metallicity problem, depends on both background control and on the amount of running time dedicated to a CNO search. If the SNO+ collaboration commit most of their experimental run-time to a search for neutrino-less double-beta decay~\cite{Andringa:2015tza} then it is possible (though perhaps unlikely) they will be overtaken by Argo, or potentially the Jinping Neutrino Experiment~\cite{JinpingNeutrinoExperimentgroup:2016nol}, despite the fact that SNO+ is in a more advanced stage of development. As an estimate, if SNO+ finishes its neutrino-less double-beta decay search in 2022, then a CNO flux measurement may be possible by around 2025 to 2027, but possibly later. This relies crucially on the estimates of the $^{210}$Bi rate in ref.~\cite{Andringa:2015tza} being correct. The Borexino experiment has the advantage that it is already running, and may be the first to exclude a zero CNO flux, provided its backgrounds are kept under control.  
Importantly, this also means that its background rates will be the most realistic of
the experiments considered in this work, in
contrast to the projections of SNO+ and
Argo, and so for a fair comparison this fact
must be taken into account.

As a final point, we note that it is possible that new technologies may allow the CNO flux to be measured by electron-recoil experiments sooner, in particular the development of experiments which can detect both scintillation and Cherenkov light, such as THEIA~\cite{Gann:2015fba,Alonso:2014fwf,Caravaca:2016fjg}. This would mean that the direction of the recoiling electrons could be measured in addition to their energies, which would break the degeneracy between solar neutrinos and background such as $^{210}$Bi.

\section{Nuclear-recoil experiments searching for CNO neutrinos}
\begin{figure}[bt]
\centering
\includegraphics[width=0.7\textwidth]{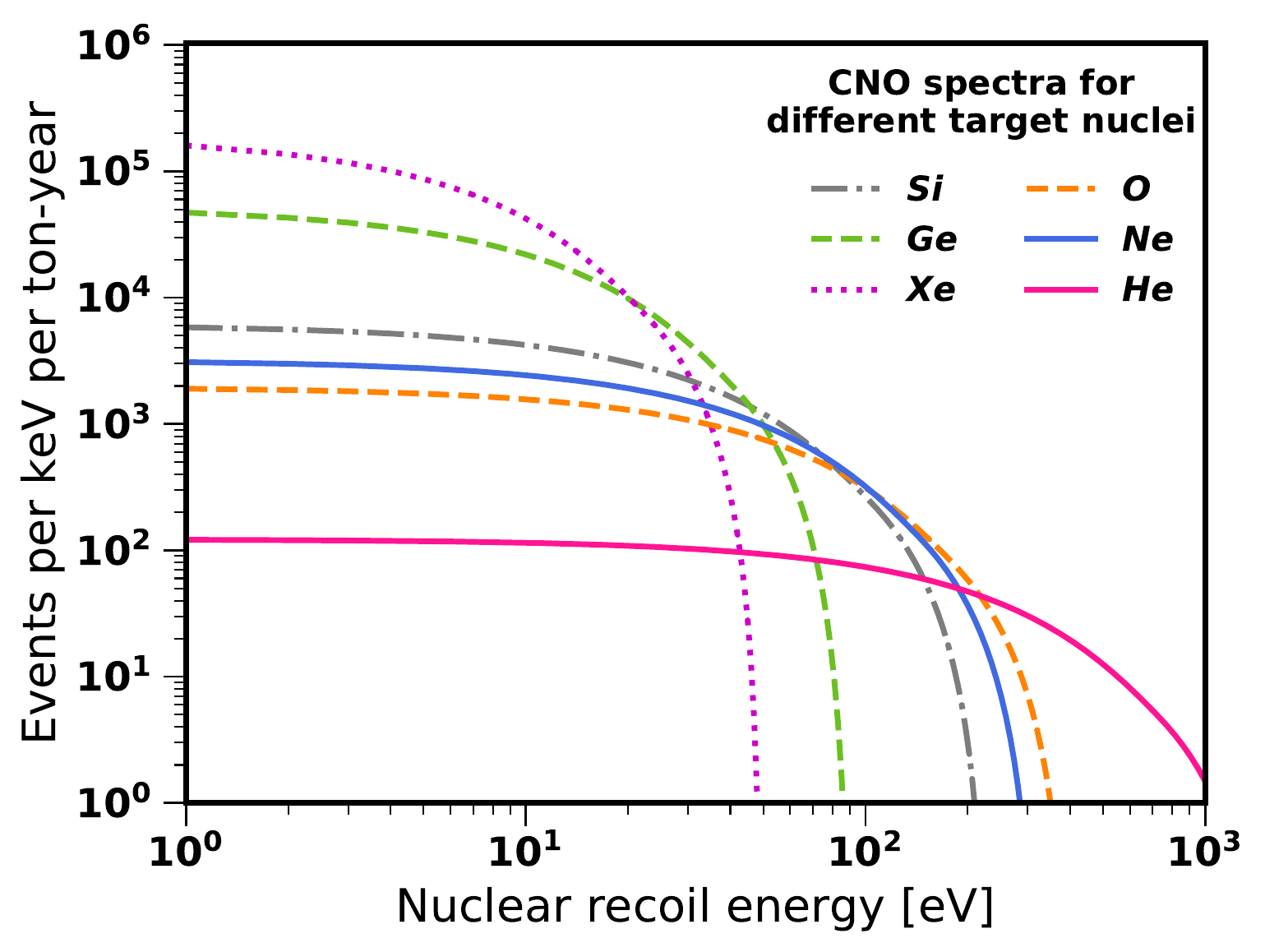}
\caption{The expected spectrum of nuclear-recoils from elastic scattering by CNO neutrinos in the high-metallicity case, for various different target nuclei.}
\label{fig:CNO_NR}
\end{figure}

In the previous section we considered experiments looking for CNO neutrinos through their scattering with electrons, now we turn to nuclear-recoil searches. For the case of CNO neutrinos scattering with nuclei the energy range of interest is below a  few hundred eV, with the exact value depending on the particular target nucleus~\cite{Strigari:2016ztv} i.e. targets with heavier nuclei require lower thresholds to observe CNO neutrinos, as shown in figure~\ref{fig:CNO_NR}. Such low-energy recoils are difficult to observe, limiting the sensitivity of these searches. However nuclear-recoil searches have the advantage of lower backgrounds, and a larger cross section of interaction between neutrinos and nuclei (compared with electrons), arising from coherent enhancement by approximately the square of the number of neutrons in the target nucleus. This also means that target exposures (effective mass times experiment live time) can in principle be much lower. Hence, in contrast to the previous section, our focus here will not be on background control, but instead we are predominantly concerned with finding the required combination of target nucleus, low-energy threshold and exposure in order to measure the CNO neutrino flux with precision.

Current nuclear-recoil experiments focus on searching for dark matter, but they also work well as neutrino detectors~\cite{Cerdeno:2016sfi,Lang:2016zhv,Strigari:2016ztv,Billard:2013qya,Billard:2014yka}. 
The energy thresholds of the most successful dark matter direct detection experiments are currently too high for a CNO neutrino search, for example the liquid-xenon-based LZ and XENON1T experiments have thresholds around a keV, meaning they will not see any CNO neutrinos~\cite{Akerib:2016vxi,Aprile:2017iyp,Mount:2017qzi}. However there has been much recent progress on the development of experiments with lower thresholds. For example, the CRESST-III experiment which looks for small temperature changes of a cryogenically-cooled CaWO$_4$ crystal, had a low-energy threshold conservatively set
at 100 eV for their most recent analysis with $2.39$~kg-days of data~\cite{Petricca:2017zdp}. In addition, the same collaboration has developed the $\nu$-cleus detector with a $20$~eV threshold using a $0.5$g Al$_2$O$_3$ crystal target~\cite{Angloher:2017sxg}. There has also been progress using germanium-based experiments such as SuperCDMS, which has achieved energy thresholds as small as 75eV and 56eV in their most recent runs with an exposure around 70kg-days~\cite{Agnese:2017jvy}. This has been possible thanks to a special (high-voltage) operation mode and new efforts in understanding and reducing the overall background rate.

The next stage in this experiment, SuperCDMS in SNOLAB, could go as low as a 40eV threshold with an exposure in high-voltage detectors around $44$kg-yr for germanium and $10$~kg-yr for silicon~\cite{Agnese:2016cpb}. In addition, the next phase of the EDELWEISS experiment, which also uses a germanium target, could have a mass as large as 100kg~\cite{Arnaud:2017usi}.  Finally, the NEWS-G experiment, based on the technology of gaseous spherical detectors, had a 720eV threshold in their recent run with  a 9.7kg-days exposure using neon and CH$_4$ targets~\cite{Arnaud:2017bjh}. In this kind of detector, the low energy threshold is limited only by the mean ionization energy of the gas mixture, which, depending on the specific target, can be as low as a few eV \cite{Bougamont:2011zz,Savvidis:2016wei}.

In the rest of this section we will focus on making projections for future versions of such searches, since at present their low target masses, much smaller than for the electron-recoil experiments, will not provide enough data for a precise measurement of the CNO flux. We will determine exactly what exposures will be needed, for a given low-energy threshold and target nucleus, for a positive detection of the CNO flux.

\subsection{Statistical procedure}

In order to do so, we test the ability of an experiment with a given target nucleus, recoil detection threshold and exposure to discriminate the CNO flux from the pp, pep, $^8$B and $^7$Be neutrino fluxes originating in the pp chain. Thus, we
\begin{enumerate}
\item randomly generate events based on the total expected spectrum, from both the proton-proton-chain and CNO. For each pseudoexperiment, the normalizations of the proton-proton-chain fluxes are randomly obtained, within measured uncertainties \cite{Bergstrom:2016cbh}\footnote{11\%, 8.5\%,15\% and 3\% for pp, $^8$B, pep and $^7$Be, respectively.}. The CNO flux is fixed to the sum of the $^{13}$N and $^{15}$O fluxes. 
\item We construct an extended unbinned likelihood: 
\begin{equation}
L = e^{-N_{\mathrm{exp}}}\prod_{i}^{N_{\mathrm{obs}}}\left[\sum_c \phi_{0,c}\frac{dR}{dE_R}(E_i)\right] P(\phi_{0,c}),
\label{eq:LLNR}
\end{equation}
where $\frac{dR}{dE_R}$ is the expected differential neutrino-nucleus scattering rate, $N_{\mathrm{obs}}$ is the number of ``observed'' events in the sample, and $P(\phi_{0,c})$ is a Gaussian prior on the neutrino fluxes, with mean and width based on the currently measured fluxes. The index c runs over the components $c = \{$pp, pep, $^8$B, $^7$Be, ($^{13}$N + $^{15}$O)  $\}$. The total solar fluxes $\phi_{0,c}$ are normalized such that the total number of expected events is:
\begin{equation}
N_{\mathrm{exp}} = \sum_c N_{\mathrm{exp},c}= \sum_c \phi_{0,c'}\int_{E_{th}}^{E_{\mathrm{max}}}\frac{dR}{dE_R}dE_R
\end{equation}
We do not consider background events, which implies a certain level of discrimination between electronic and nuclear recoils.

This is because solar neutrinos from other populations, especially $pp$, will induce electron recoils in the energy range relevant for a nuclear-recoil search for CNO neutrinos.  Despite $pp$ neutrinos being much more abundant, coherent scattering with nuclei benefits from a larger cross section than for electronic scattering, including a factor of the nucleon number squared.  
With this consideration, the rate expected from $pp$ neutrino-induced electron scattering is smaller than the one for CNO neutrino nuclear recoils, by approximately a factor 10 in xenon for example, in the region-of-interest~\cite{Cerdeno:2016sfi,Akerib:2018lyp}. 
Thus, although a certain level of discrimination between electron and nuclear recoils would be desirable, it is only really important for the lightest nuclei, where one would preferably have around $90\%$ electron-recoil rejection or better.

\item The likelihood in equation \eqref{eq:LLNR} is then maximised setting $\phi_{0,\mathrm{CNO}} = 0$, then again for  allowing the total CNO flux to vary freely. The quantity $ts = 2(\log(L_{\mathrm{CNO}}) - \log(L_{\mathrm{CNO}=0}))$ is constructed. We count an experiment as successful when $ts > 3.84$, corresponding to a 95\% CL detection ($ts$ is indeed distributed as a chi-squared with one degree of freedom). 
\item Steps 1-3 are repeated 5000 times for each point in exposure-threshold parameter space. 
\end{enumerate}

\begin{figure}[t]
\centering
\includegraphics[width=0.9\textwidth]{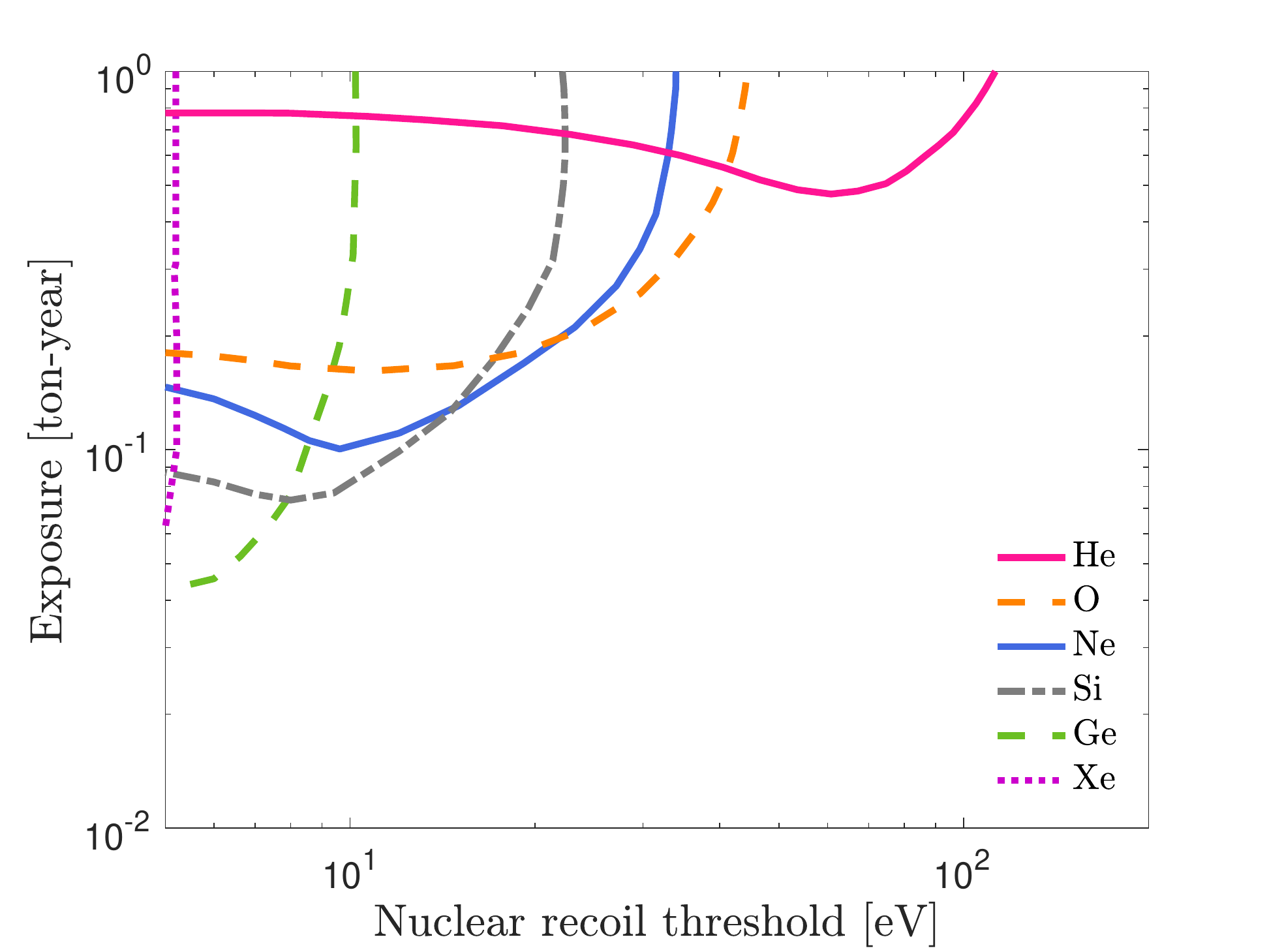}
\caption{Required exposure versus threshold to discover the CNO neutrino flux via coherent neutrino-nucleus interaction at 95\% CL, in the high metallicity scenario. For each target element all parameters above the lines will lead to a CNO detection. In the low-metallicity case, we are unable to significantly distinguish the CNO flux from the other solar neutrino components.}
\label{fig:NR_results_high}
\end{figure}

\subsection{Results}

Fig.~\ref{fig:NR_results_high} shows the result of this procedure, for nuclear recoil experiments using helium, oxygen, neon, silicon, germanium and xenon in the high-metallicity scenario. The contours represent the line above which 90\% of the pseudoexperiments are able to see the CNO flux at 95\% CL or more. 
As expected, lighter targets are able to measure the CNO flux with a higher threshold, as larger momenta can be transferred from the neutrinos. The drawback is a suppression in event rate due to the smaller $N^2$ enhancement, where $N$ is the number of neutrons.

Figure~\ref{fig:NR_results_high} allows us to assess the suitability of the different experimental techniques to look for CNO neutrinos. For example, despite their large projected exposures, liquid xenon detectors would require a threshold energy smaller than 5eV, which is three orders of magnitude smaller than their current values. It is not clear if this is physically possible, see e.g. ref. \cite{Baldini:2004td}.
Conversely, the threshold needed for gaseous spherical detectors with light targets seems within the reach of future experiments, however, the required exposures are still very far from those projected in future detectors, and it is not clear whether discrimination from electron recoils is possible. The situation is similar for experiments with oxygen-based crystal targets, such as CRESST and $\nu$-cleus~\cite{Angloher:2017sxg}.

Of all the nuclear-recoil technologies, low-temperature solid state detectors appear to be best suited for the observation of CNO neutrinos. Indeed, the exposure needed for germanium is just a factor of three larger than what SuperCDMS SNOLAB will achieve, although the threshold would need to be reduced to $\sim 10$eV. Likewise, silicon would require a threshold of approximately 30eV, smaller than the projected 70eV of SuperCDMS SNOLAB, and the exposure needed is approximately ten times larger.
In principle, these thresholds could be reduced since the required energy to produce a single electron-hole ($e-h$) pair in germanium and silicon is of the order of 3eV and 3.6eV, respectively. Although this is not within the reach of current technology, future improvements could make this possible.

For detection of CNO neutrinos in the low-metallicity scenario, the prospects of direct detection experiments are less promising. The $^{13}$N and $^{15}$O fluxes decrease substantially and the resulting spectrum blends in to that of the pep neutrinos. As a result, much larger exposures are needed, as well as much lower thresholds. As an example, a silicon experiment would require a threshold of approx 10~eV and a minimum exposure of $0.5$~ton-yr to start observing these events, whereas a germanium-based experiment would require a threshold close to the minimum energy to excite an $e-h$ pair.

\section{Conclusion}
There is no consistent model of the Sun which explains both helioseismological data and measurements of the solar metallicity using spectroscopic data, leading to the so-called ``solar metallicity problem''~\cite{Asplund:2009fu,2009ASPC..416..193B}.
The aim of this paper has been to work out when experiments will measure the CNO neutrino flux with enough precision to help solve this problem, and the challenges this involves. We have studied both nuclear recoil experiments, which need to be sensitive to very low-energy recoils but have small backgrounds, and electronic recoil experiments, which generally have larger target masses but also higher background rates. We determine whether such nuclear-recoil technologies, which are important to the search for light dark matter, can catch up with the electronic-recoil experiments.

For experiments searching for CNO neutrinos using electron-recoils, we have made projections with an MCMC analysis, to compare how precisely they will be able to measure the CNO flux, given various assumptions about the backgrounds and how well their rates will be measured. Figure~\ref{fig:borexino_mcmc} shows our results for Borexino, figure~\ref{fig:sno_plus} for SNO+ and figures~\ref{fig:argon_GS_1000tonyr} and \ref{fig:argon_SNOlab_1000tonyr} for a liquid argon experiment, such as Argo~\cite{Aalseth:2017fik,Franco:2015pha}, in Gran Sasso and SNOLAB respectively. Each two-dimensional panel shows the range of parameters which provide a good fit to simulated data, highlighting the degeneracies between parameters which lead to systematic uncertainties. For Borexino and SNO+ the main degeneracy is between the CNO flux and the $^{210}$Bi background, while for argon experiments the CNO flux is degenerate with the background from the decay of the $^{222}$Rn daughters. Any technology which could break this degeneracy would significantly improve the accuracy of a CNO flux measurement (see e.g. refs.~\cite{Gann:2015fba,Alonso:2014fwf,Caravaca:2016fjg}). In all cases the CNO flux is strongly degenerate with the pep neutrino flux.

Since we have performed our own analysis, we are able to compare our projected sensitivities for each experiment. Our comparison is shown in figure~\ref{fig:ER_comparison}. For each experimental run we have chosen an optimistic and pessimistic prior set, with the former imposing strong constraints on the sizes of key backgrounds and thereby reducing systematic uncertainties, while in the latter case we use only weak constraints. The comparison between these different assumptions makes it clear that controlling key systematics is just as vital as obtaining more data.

For experiments looking for CNO neutrinos through their scattering with nuclei we  focused on future searches, and determined the required low-energy threshold and exposure for different target nuclei, in order to measure the CNO neutrino flux precisely. Our results for nuclear-recoils are shown in figure~\ref{fig:NR_results_high}.

Our analysis highlights several ways in which the CNO flux may be measured to the precision needed to separate the two solar metallicity scenarios, which we list below:
\begin{itemize}
\item If Argo is built in Gran Sasso lab with a $100$~ton fiducial mass then it will measure the CNO flux to the required precision after ten years, provided that the $^{222}$Rn-chain background is measured to a accuracy of $10\%$ or better.
\item Building Argo in SNOLAB or Jinping, where the cosmogenic backgrounds will be smaller, should lead to an accurate CNO flux measurement in only $5$~years, assuming a $100$~ton fiducial mass. In all cases the argon experiments rely on their extremely good projected energy resolution.
\item SNO+ will measure the CNO flux in $5$~years provided that the $^{210}$Bi rate is known to at least $1\%$ accuracy. However this is only if it is kept running in its liquid scintillator mode, and not while doped with $^{130}$Te for the neutrino-less double-beta decay search. It is likely the best candidate to see CNO neutrinos, with a potential for detection between 2025 to 2027 if its double-beta search ends in 2022, but only if the $^{210}$Bi rate meets projections to be at least as low as that already achieved in Borexino~\cite{Andringa:2015tza}.
\item Borexino will set strong limits on the CNO neutrino flux, and could obtain a good measurement after ten years, provided it can measure the $^{210}$Bi rate to an accuracy of better than $0.5$~counts per day. It is the most sensitive of the experiments to systematics from backgrounds, and so an accurate measurement of $^{210}$Bi is particularly important. Crucially though, it is the only experiment we have considered which is currently running, and so has the most realistic background assumptions.
\item Nuclear recoils of CNO neutrinos could also be within the reach of dark matter experiments, although these would require very low-thresholds and large exposures. Low-temperature solid state detectors seem the best alternative to confirm or rule-out the high-metallicity scenario, but the technology will have to improve to be able to lower the experimental threshold down to 10eV for germanium or 30eV for silicon. Probing the low-metallicity scenario is much more challenging, as lower thresholds and larger exposures are needed. Such an experiment would also revolutionize constraints on light dark matter by many orders of magnitude~\cite{Agnese:2017jvy}.
\end{itemize}

It is clear is that, despite the challenges, neutrino experiments will be able to contribute significantly to the solar metallicity problem in the near future. A full solution will likely also need us to understand why helioseismology disagrees with the standard solar model. Which experiment gets there first will depend on the amount of time dedicated to a CNO neutrino search and crucially, how well the backgrounds can be controlled for electron-recoil searches, and how low the energy threshold will be for nuclear-recoil searches.

\acknowledgments
We thank Davide Franco, Gabriel Orebi Gann, Gilles Gervais, El\'ias L\'opez-Asamar, Valentina Lozza, Ryan Martin and Oleg Smirnov.
DGC acknowledges support from STFC.
JD and MF are funded by the European Research Council through the project DARKHORIZONS under the European Union's Horizon 2020 program (ERC Grant Agreement no.648680).  The work of MF was also supported partly by the STFC Grant  ST/L000326/1 and this project has benefited from a Durham IPPP Fellowship.   

\bibliographystyle{JHEP}

\providecommand{\href}[2]{#2}\begingroup\raggedright\endgroup

\end{document}